\documentclass[]{spie}
\usepackage[utf8]{inputenc}
\usepackage[T1]{fontenc}
\usepackage{amsmath, amssymb}
\usepackage{graphicx}
\usepackage{cite}
\usepackage{hyperref}
\usepackage{authblk}
\usepackage{graphicx}
\usepackage{subcaption}
\usepackage{ragged2e} 
\usepackage{tabularx} 
\usepackage{booktabs} 
\usepackage{caption} 
\usepackage{multirow} 
\usepackage{multicol} 
\usepackage{makecell}
\usepackage{float}
\usepackage{comment}

\title{Characterizing the performance of two C-RED ONE cameras for implementation in RISTRETTO and SAXO+ projects}
\author[1]{Muskan Shinde}
\author[1]{Jana Anouk Baron}
\author[1]{Nicolas Blind}
\author[1]{Janis Hagelberg}
\author[1]{Christophe Lovis}
\author[1]{François Wildi}
\author[1]{Damien Ségransan}
\affil[1]{D\'epartement d'Astronomie, Universit\'e de Gen\`eve, Chemin Pegasi 51, CH-1290 Versoix, Switzerland}
\date{} 
\begin{document}
\maketitle

\begin{abstract}
\noindent
In the near-infrared wavelength regime, atmospheric turbulence fluctuates at a scale of a few milliseconds, and its precise control requires the use of extreme adaptive optics (XAO) systems equipped with fast and sensitive detectors operating at kHz speeds. The C-RED One cameras developed by First Light Imaging (FLI), based on SAPHIRA detectors made of HgCdTe e-APD array sensitive to 0.8-2.5 $\mu$m light, featuring a 320×256 pixels with 24 $\mu$m pitch, offering sub-electron readout noise and the ability to read subarrays, at frame-rates of up to few 10-kHz, are state-of-the-art for XAO wavefront sensing. The Observatory of Geneva purchased two C-RED One cameras identified as necessary for RISTRETTO (a proposed high-contrast high-resolution spectrograph for the VLT) and SAXO+ (an upgrade of the VLT/SPHERE XAO system) projects. We present a comprehensive characterization and comparative analysis of both the cameras. We present test results examining key noise contributors, including readout noise, detector bias, etc. And we also study their temporal variability. Additionally, we assess the conversion gain and the avalanche gain calibration of the detector. We also study the evolution some of these parameters over time.
\end{abstract}


\section{Introduction}
\noindent
In the near-infrared wavelength regime, atmospheric turbulence fluctuates at a scale of a few milliseconds, and its precise control requires the use of extreme adaptive optics (XAO) systems equipped with fast and sensitive detectors operating at kilohertz (kHz) speeds. Charge-coupled devices (CCDs) cannot operate in this regime; therefore, complementary metal-oxide-semiconductor (CMOS)-based detectors are used. However, the ability to use these CMOS detectors is limited due to the readout noise of the detector. The longstanding challenge in the near-infrared has been the limiting factor of detector readout noise. To overcome the detector noise, a good approach is to amplify the photoelectron signal through an avalanche process. Therefore, electron-amplifying avalanche photodiodes (eAPDs) are of great interest.
One of the best materials for building infrared eAPD arrays is mercury cadmium telluride (HgCdTe) because it is a direct semiconductor. The APD gain process in HgCdTe is almost noiseless \cite{finger2014saphira}. The most promising features of HgCdTe are the uniform APD gain, which increases exponentially with the applied bias voltage, and the absence of avalanche breakdown \cite{finger2010development}.

The European Southern Observatory (ESO) initiated a program with SELEX-Galileo Infrared Ltd to develop near-infrared electron avalanche photodiode arrays (eAPDs) for wavefront sensing and fringe tracking. They developed the Selex Advanced Photodiode HgCdTe Infrared Array (SAPHIRA) detector, which offers sub-electron readout noise and the ability to read subarrays for increased frame rates, making them well-suited for wavefront sensing. The SAPHIRA detector utilizes the HgCdTe APD properties and has a user-adjustable avalanche gain (eAPD-gain) that amplifies the signal from photons but not the readout noise. At high APD gain, this reduces the readout noise well below the sub-electron level at frame rates of 1 kHz. The current growth technology used is metal-organic vapor phase epitaxy (MOVPE). This technology provides more flexibility in the design of diode structures, allowing the creation of heterojunctions with different bandgap properties between the absorption region and the multiplication region. The shift to MOVPE has resulted in a dramatic improvement in cosmetic quality, with 99.97\% operable pixels at an operating temperature of 85K \cite{greffe2016c}. First Light Imaging(FLI) transformed the SAPHIRA detectors into the commercial product named the C-RED One camera, which promises less than 1 electron readout noise and a full-frame readout speed of up to 3500 frames per second.

In the context of developing XAO in the near-infrared, in 2023 we purchased we purchased two C-RED ONE cameras, designating one for integration into the RISTRETTO instrument \cite{chazelas2020ristretto, lovis2022ristretto}, a proposed high-contrast high-resolution spectrograph for the VLT, and the other for the SAXO+ instrument \cite{stadler2022saxo+}, an upgrade of the VLT/SPHERE XAO system. A description of the C-RED One camera is provided in Section 2. The detector characterization, including readout noise measurement, system gain calculation through the photon transfer curve (PTC), effective gain calculation, and full well capacity analysis, is explained in Section 3. The results and conclusions are presented in Section 4.

\section{The CRED One camera}
\noindent
The C-RED One camera incorporates a SAPHIRA detector and is capable of capturing up to 3500 full frames per second with sub-electron readout noise. Although the SAPHIRA detector can be used in the 0.8–2.5 $\mu m$ wavelength range, this camera is designed to include four long-wavelength blocking filters that limit sensitivity to 1.75 $\mu m$. This step is taken to filter out unwanted background flux outside the J and H-band wavelength window, which would otherwise be amplified at low gain operation \cite{gach2016c}. The sensor is placed in a sealed vacuum environment and cooled to a temperature of 80 K using an integrated pulse tube. The pixel format is 320x256 pixels. The array has 32 parallel video outputs, organized as 32 sequential pixels in a row. The 32 outputs are arranged in such a way that the full multiplex advantage is available even for small sub-windows. Non-destructive readout schemes with subpixel sampling are possible. The camera can operate at a frame rate of 3.5 kHz for full frames. The frame rate can be further increased by using sub-windowing. For a 32x32 sub-window, the maximum frame rate can go up to 100 kHz. The characteristics of the detector are listed in Table~\ref{tab:camera_specs}.

\vspace{12pt} 

\noindent
\textbf{Specifications of the CRED-One camera} \\
\noindent
Detector type: SAPHIRA \\
Resolution: 320x256 pixels \\
Pixel pitch: 24 $\mu m$ \\
Quantization: 16 bit \\
Quantum Efficiency: $>$ 60\% from 1.1 to 2.4 $\mu m$ \\
Maximum speed: 3500 fps, corresponding to 285 $\mu s$ exposure time \\
Minimum speed: 0.02 fps, corresponding to 50 $s$ exposure time \\
Wavelength Range: 0.8 to 3.5 $\mu m$

\begin{table}[ht]
\centering
\caption{Detector measurements given in FLI user manual \\}
\vspace{1em}
\begin{tabular}{|l|l|l|}
\hline
\textbf{Specification} & \textbf{Unit} & \textbf{Measurement} \\ \hline
Mean readout noise at 3500 fps & $e$ & $<0.5$ \\ \hline
Dark signal at 3500 fps at 80k & $e/\text{pixel}/\text{frame}$ & $0.03$ \\ \hline
Detector operating temperature & K & $80$ \\ \hline
Peak quantum Efficiency at 1550 nm & \% & $70$ \\ \hline
Image full well capacity at gain 1 & $e$ & $60,000$ \\ \hline
Latency between exposure and first pixel availability & $\mu s$ & $2$ \\ \hline
\end{tabular}
\label{tab:camera_specs}
\end{table}

\section{Camera characterization}
\noindent
The camera offers three different readout modes, namely the Global reset single (GRS), the Correlated double sampling (CDS) and the Multiple non-destructive reads (MND). The readout mode we use is the CDS mode. In this mode, the entire frame is read immediately after a frame demand then reset and automatically read again. The result of the CDS processing is the difference between those two frames and gives a flow in e-/frame. The maximum speed is twice lower than the maximum frame rate in single read mode. All the characterization tests we performed are in the CDS mode. We compare our tests with the tests report of each of the two camera provided by FLI.

\subsection{Subwindowing}

The camera offers the possibility of subwindowing. This feature allows the selection of one or multiple regions of interest on the detector. Only the selected region is read out, and in the case of multiple regions, their intersection is read out. The ability to read only a selected region enables a higher frame rate. In the CDS mode, the maximum frame rate is 1738 Hz for a full 320x256 frame, which can increase up to 44667 Hz for a frame size of 32x32, as detailed in Table~\ref{tab:height_width}.

\begin{table}[ht]
\caption{In the CDS mode, the maximum framerate achievable in Hz for different subwindow sizes.}
\vspace{1em}
\centering
\begin{tabular}{|c|c|c|c|c|c|}
\hline
\makecell{height/width \\ (pixels)} & 320 & 256 & 128 & 64 & 32 \\ \hline
256                                & 1738 & 2117 & 3757 & 6132 & 8965 \\ \hline
128                                & 3387 & 4114 & 7204 & 11538 & 16502 \\ \hline
64                                 & 6446 & 7785 & 13311 & 20637 & 28470 \\ \hline
32                                 & 11754 & 14056 & 23103 & 34068 & 44667 \\ \hline
\end{tabular}
\label{tab:height_width}
\end{table}

\subsection{Detector bias}
\begin{figure}[H]
    \centering
    \begin{subfigure}[b]{0.45\textwidth}
        \centering
        \includegraphics[width=\textwidth]{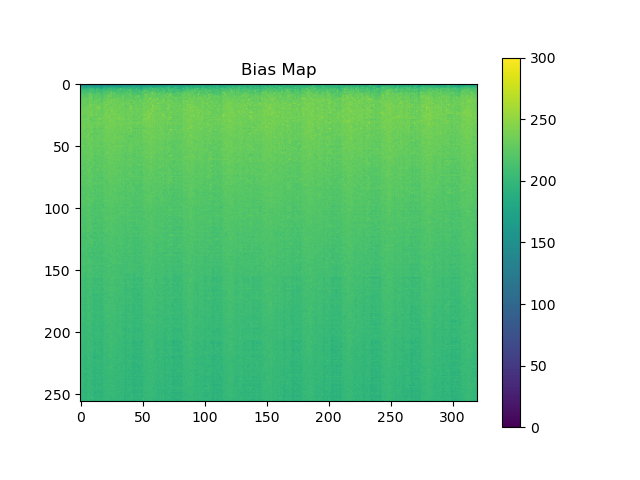} 
        \caption{Camera 1}
        \label{subfig:camera1}
    \end{subfigure}
    \hfill
    \begin{subfigure}[b]{0.45\textwidth}
        \centering
        \includegraphics[width=\textwidth]{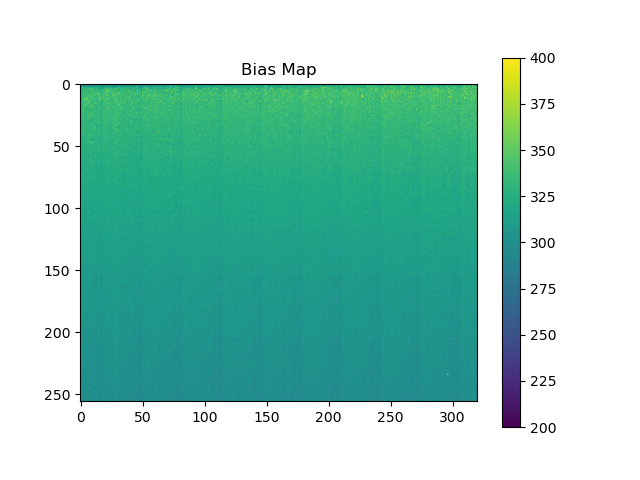} 
        \caption{Camera 2}
        \label{subfig:camera2}
    \end{subfigure}
    \caption{The bias map in the CDS mode produced by averaging the 2000 frames. The left image corresponds to camera 1, while the right image corresponds to camera 2. The colorbar represents the flux level in ADU.}
    \label{fig:bias_map}
\end{figure}

To examine the bias level of the camera, dark frames were acquired while the camera cap was closed. The frames were taken at an eAPD-gain of 1 and at minimal exposure time, ensuring that the dark current was sufficiently low, thus capturing the true bias level of the camera. A data cube of 2000 frames was taken and temporally averaged to generate a bias map, which is shown in Figure~\ref{fig:bias_map}. This map reveals a recurring bias pattern observed within columns of pixels along the readout direction. For camera 1, the measured median bias value is 214 ADU, while for camera 2 it is 314 ADU.

\begin{figure}[H]
    \centering
    \begin{subfigure}[b]{0.4\textwidth}
        \centering
        \includegraphics[width=\textwidth]{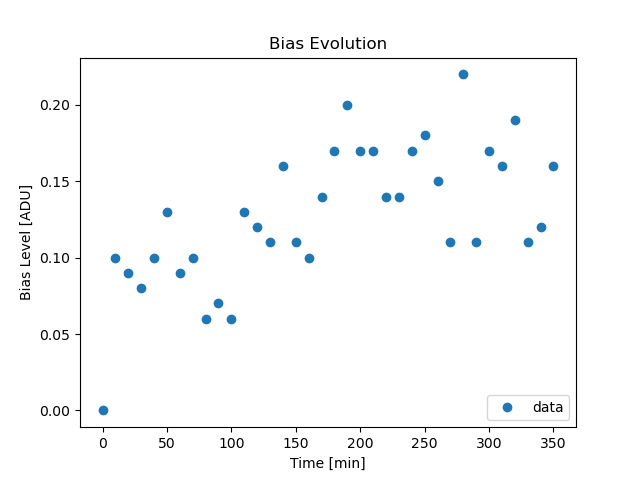} 
        \caption{Camera 1}
        \label{subfig:camera1}
    \end{subfigure}
    \hfill
    \begin{subfigure}[b]{0.4\textwidth}
        \centering
        \includegraphics[width=\textwidth]{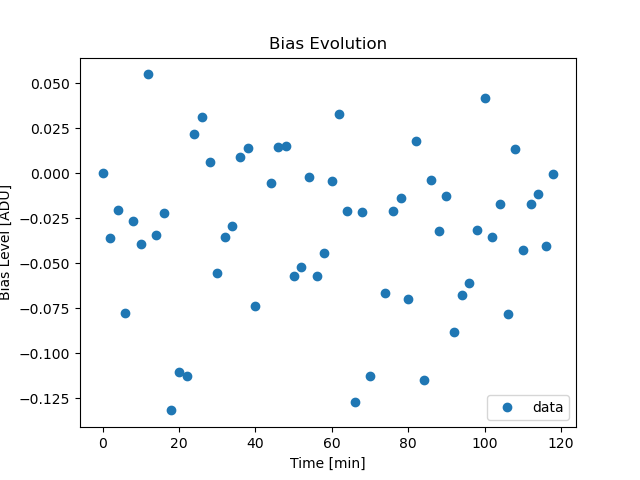} 
        \caption{Camera 2}
        \label{subfig:camera2}
    \end{subfigure}
    \caption{The variation of the bias level over an approximate 8-hour period is observed. At intervals of 10 minutes, 100 bias frames are captured at the maximum framerate and subsequently averaged to generate a bias map. To better visualize the bias evolution over time, each pixel has the initial reference bias subtracted from it. The left image corresponds to camera 1, while the right image corresponds to camera 2.}
    \label{fig:bias_evolution}
\end{figure}

In Figure~\ref{fig:bias_evolution}, the change of the bias level over an 8-hour period is displayed. To better visualize the bias evolution over time, the initial reference bias is subtracted from each pixel. The bias demonstrates variations within a range of approximately 0.3 ADU. We conducted an analysis to assess the impact of bias level variation on the wavefront sensor's performance. The results indicated that the variation has an insignificant effect.

\subsection{Flat field}
\begin{figure}[htbp]
    \centering
    \includegraphics[width=0.7\textwidth]{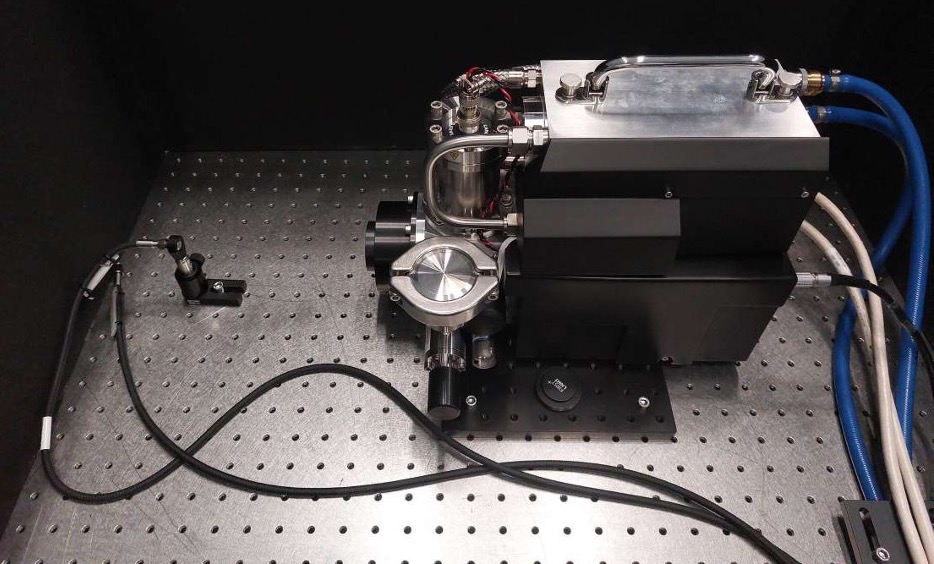} 
    \vspace{1em}
    \caption{The optical bench setup designed to achieve uniform illumination on the camera. An optical fiber, coming from a highly stable light source, is securely positioned in front of the camera at a distance ensuring the attainment of uniform illumination. To prevent contamination from stray light, the entire arrangement is enclosed within a black box.}
    \label{fig:optical_bench}
\end{figure}

For the characterization of the camera, flat field frames were acquired. The flat field frames were obtained by illuminating the detector with an infrared laser source. The light from the light was injected into an optical fiber which is fixed in front of the camera at a distance such that uniform illumination is achieved on the camera. To prevent stray light contamination, the entire setup was enclosed within a black box. The picture of the setup is shown in Figure \ref{fig:optical_bench}. The optical setup was not designed for precise flat field estimation but rather to provide close to uniform illumination for characterization tests. For the purpose of these tests, we would be referring to the image obtained in this setup as “flat fields”.

\begin{figure}[H]
    \centering
    \begin{subfigure}[b]{0.45\textwidth}
        \centering
        \includegraphics[width=\textwidth]{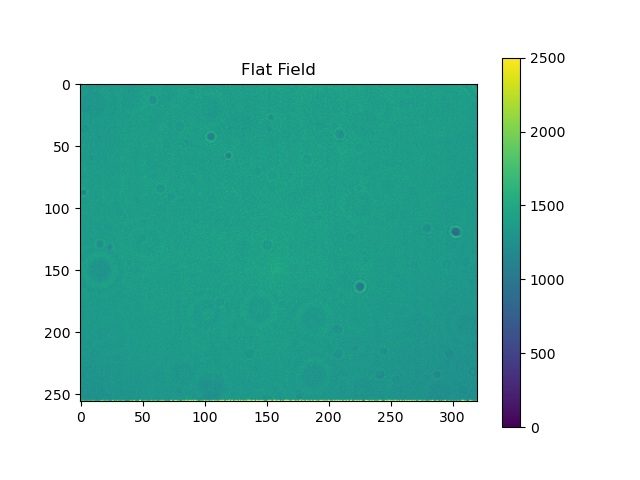} 
        \caption{Camera 1}
        \label{subfig:camera1}
    \end{subfigure}
    \hfill
    \begin{subfigure}[b]{0.45\textwidth}
        \centering
        \includegraphics[width=\textwidth]{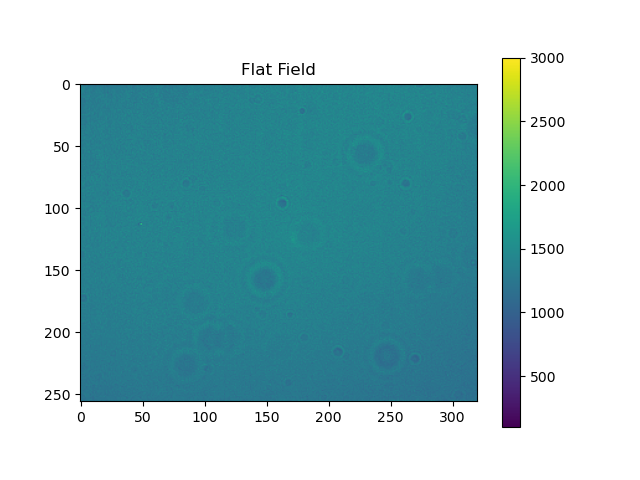} 
        \caption{Camera 2}
        \label{subfig:camera2}
    \end{subfigure}
    \caption{Images from the camera obtained with uniform illumination, taken at the frame rate of 500 fps. The left image corresponds to camera 1, while the right image corresponds to camera 2. The colorbar represents the flux level in ADU.}
    \label{fig:uniform_illumination}
\end{figure}

\vspace{1cm}
Figure \ref{fig:uniform_illumination} shows an image captured within this setup at fps of 500 Hz. Numerous rings of varying sizes and intensities can be seen in the images. These rings are the result of diffraction caused by dust particles present on different layers of filters preceding the detector. Notably, the location and distribution of these rings vary based on the light source's distance and angle. This indicates that the dust is on the filters in front of the detector and not on the detector itself. However, the presence of dust does not affect the performance of the camera.

\subsection{Defective pixels}
To identify defective pixels, 100 frames were taken at the maximum framerate of 1740 fps and at the maximum eAPD-gain level, which is 108 for camera 1 and 96 for camera. The frames are taken at maximum eAPD-gain as this is the configuration we will be in when observing on sky with this camera. We then temporally averaged the frames and obtained the median flux for each individual pixel across the full frame of the detector. Subsequently, we flagged pixels with median flux values exceeding a certain threshold (termed "hot pixels") or falling below a certain threshold (termed "dead pixels") as defective.

\begin{figure}[H]
    \centering
    \begin{subfigure}[b]{0.4\textwidth}
        \centering
        \includegraphics[width=\textwidth]{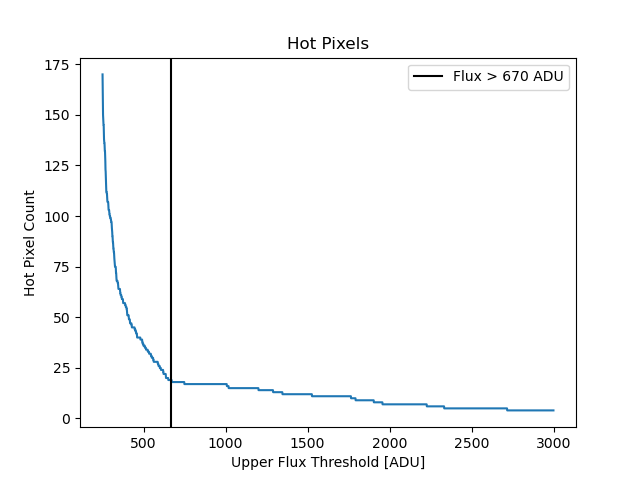}
        \caption{Camera 1 - Hot Pixels}
        \label{fig:camera1_hot}
    \end{subfigure}
    \begin{subfigure}[b]{0.4\textwidth}
        \centering
        \includegraphics[width=\textwidth]{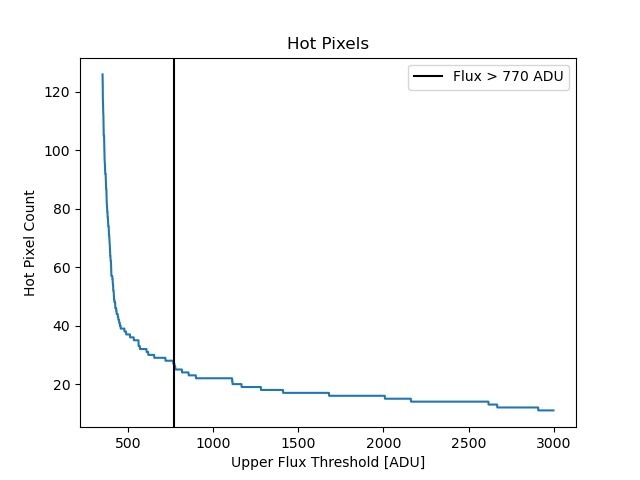}
        \caption{Camera 2 - Hot Pixels}
        \label{fig:camera2_hot}
    \end{subfigure}
    \\
    \begin{subfigure}[b]{0.4\textwidth}
        \centering
        \includegraphics[width=\textwidth]{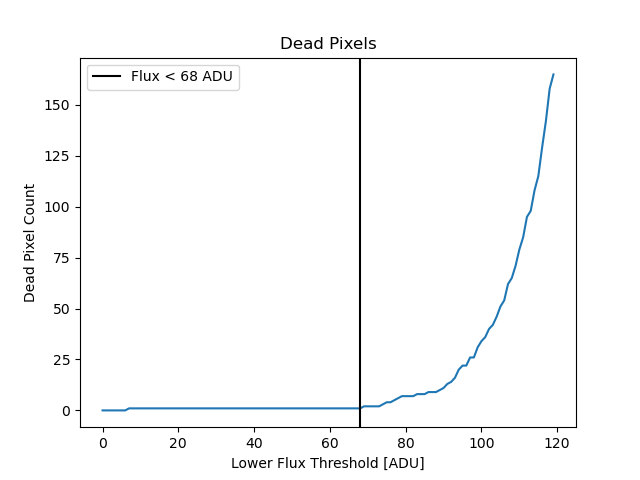}
        \caption{Camera 1 - Dead Pixels}
        \label{fig:camera1_dead}
    \end{subfigure}
    \begin{subfigure}[b]{0.4\textwidth}
        \centering
        \includegraphics[width=\textwidth]{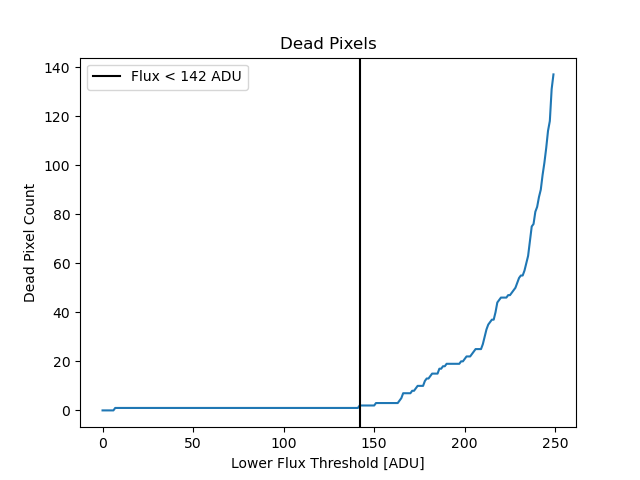}
        \caption{Camera 2 - Dead Pixels}
        \label{fig:camera2_dead}
    \end{subfigure}
    \caption{(Top) The relationship between the upper flux threshold and the count of "hot pixels" (pixels with flux values exceeding the threshold). A black vertical line indicates the threshold where the number of hot pixels starts to increase exponentially. (Bottom) The relationship between the lower flux threshold and the count of "dead pixels" (pixels with flux values falling below the threshold). A black vertical line indicates the threshold where the number of dead pixels levels off. Left plots are for camera 1, and right plots are for camera 2.}
    \label{fig:defective_pixels_cutoff}
\end{figure}

\begin{figure}[H]
    \centering
    \begin{subfigure}[b]{0.45\textwidth}
        \centering
        \includegraphics[width=\textwidth]{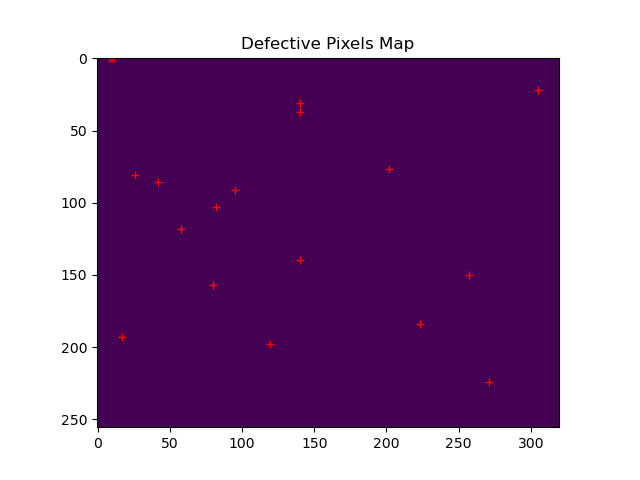} 
        \caption{Camera 1}
        \label{subfig:camera1}
    \end{subfigure}
    \hfill
    \begin{subfigure}[b]{0.45\textwidth}
        \centering
        \includegraphics[width=\textwidth]{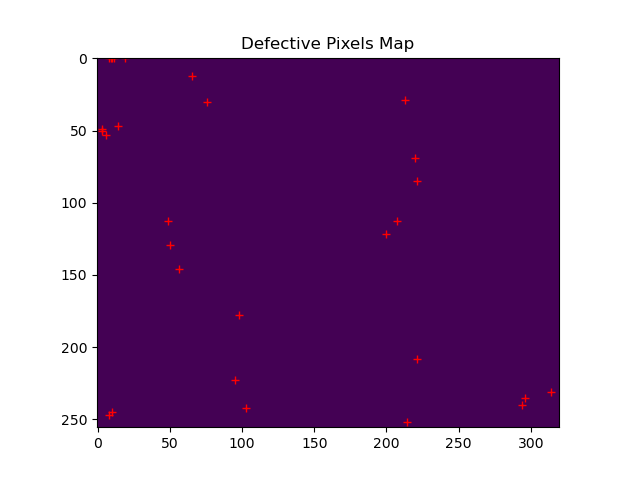}
        \caption{Camera 2}
        \label{subfig:camera2}
    \end{subfigure}
    \caption{The defective pixels, including hot and dead pixels, are marked with red crosses on the detector. (Left) For camera 1, the number of defective pixels is 20. (Right) For camera 2, the number of defective pixels is 29.}
    \label{fig:defective_pixels_map}
\end{figure}

To determine the count of defective pixels, we set the thresholds for hot and dead pixels. The relationship between the upper flux threshold and the count of hot pixels, and between the lower flux threshold and the count of dead pixels is shown in Figure~\ref{fig:defective_pixels_cutoff}. Determining accurate flux thresholds for identifying defective pixels is challenging, as the suitability of a particular threshold depends on the intended observational use of the camera. For characterization purposes, we have selected a threshold based on visual judgment. The upper threshold was defined at the point where the number of hot pixels begins to rise exponentially. Conversely, the lower threshold was established at the level where the flux associated with dead pixels becomes significantly low. For camera 1, an upper threshold of 670 ADU and a lower threshold of 68 ADU were selected. The total number of defective pixels is 20 out of the 320x256 pixels on the detector, accounting for approximately 0.024\% of the total pixels. For camera 2, an upper threshold of 770 ADU and a lower threshold of 142 ADU were found. The locations of the defective pixels identified based on these thresholds are shown in Figure~\ref{fig:defective_pixels_map}. The total number of defective pixels is 29, accounting for approximately 0.035\% of the total pixels. The number of defective pixels we have identified are more than the number reported in FLI's test reports. This discrepancy might be because our measurements were conducted at the maximum eAPD-gain, whereas FLI's measurements were presumably taken at an eAPD-gain of 1. Notably, the defective pixels we detected are spread across the detector, without any clusters. This  distribution will mean that any region of the detector can be reliably utilized for capturing the images from the wavefront sensor.

\subsection{Avalanche gain}
The e-APD technology involves applying a reverse-bias voltage across the P-N junction of each pixel. When an incident photon generates an electron, it travels through the multiplication region where the bias voltage accelerates the photoelectron along the electric field. As the electron's kinetic energy reaches a sufficient level, it ionizes a substrate atom, releasing another electron. These two electrons undergo acceleration again, generating new electrons through collision. This process is known as avalanche. By amplifying the number of electrons prior to readout, e-APD arrays multiply the signal while maintaining consistent noise. The APD gain increases exponentially with the applied bias voltage \cite{finger2010development}. The eAPD-gain is measured as a function of the APD diode bias voltage by the manufacturer and plotted on an exponential curve. Based on this curve, an electron-multiplying gain table is loaded into the camera.

We used the methodology of directly dividing the signal at eAPD-gain M by the signal at gain 1 to calculate the actual amplification of the signal at the set eAPD-gain. 

\begin{figure}[H]
    \centering
    \includegraphics[width=0.6\textwidth]{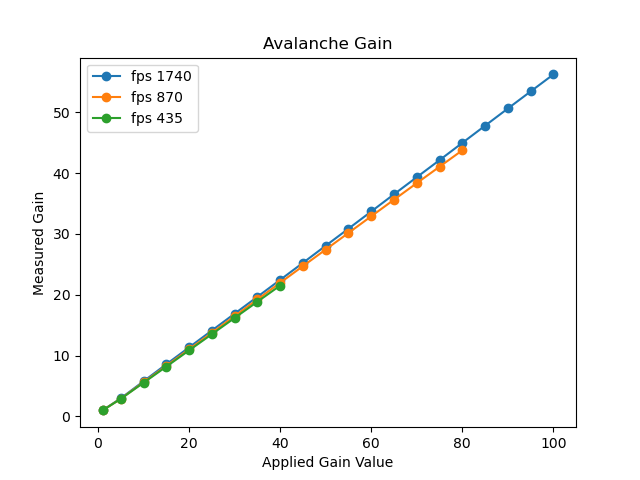} 
    \caption{The gain measured plotted against the gain applied for different frame rates: yellow - 1740 fps,  green - 870 fps, red - 435 fps. The applied gain is increased in steps of 5. The pixels included in this plot are chosen based on a 100x100 grid on the detector.}
    \label{fig:avalanche_gain_plot_different_framerates}
\end{figure}

\[
\text{Actual eAPD-gain} = \frac{S(\text{eAPD-gain } M)}{S(\text{eAPD-gain } 1)}
\]

Flat fields and dark images were acquired at various eAPD-gain levels. Using eAPD-gain 1 as a baseline, the eAPD-gain values were computed by dividing the mean flux of the dark-subtracted data frames collected at specific eAPD-gain settings by the mean flux of the dark-subtracted data frames gathered at eAPD-gain 1.

In Figure \ref{fig:avalanche_gain_plot_different_framerates}, we present a plot illustrating the relationship between the measured gain and the applied gain. For camera 1, data collection began at a frame rate of 435 fps, incrementally increasing the applied gain in steps of 5, up to a maximum gain of 40. To prevent detector saturation, the frame rate was doubled to 870 fps, and the data collection was repeated, extending the applied gain to 80. Once more, the frame rate was doubled, this time to 1740 fps, and data collection continued in 5-step increments until reaching an applied gain of 100. The measured gain exhibits a consistent linear increase with applied gain across all frame rates, with similar slopes.

\begin{figure}[]
    \centering
    \begin{subfigure}[b]{0.495\textwidth}
        \includegraphics[width=\textwidth,height=6cm]{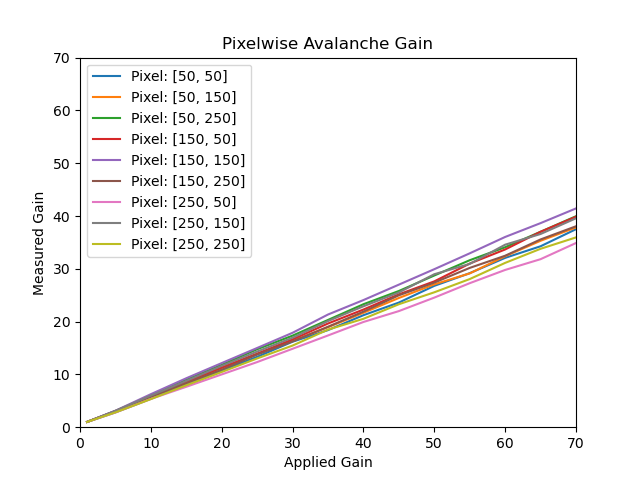} 
        \caption{Camera 1}
        \label{fig:camera1}
    \end{subfigure}
    \hfill
    \begin{subfigure}[b]{0.495\textwidth}
        \includegraphics[width=\textwidth,height=6cm]{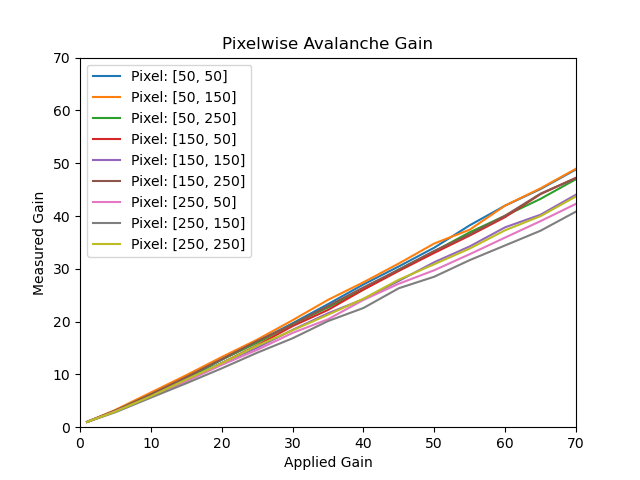} 
        \caption{Camera 2}
        \label{fig:camera2}
    \end{subfigure}
    
    \caption{The eAPD-gain measured plotted against the gain applied for a framerate of 1740 fps for specific pixels. The applied gain is increased in steps of 5. The pixels included in this plot are chosen based on a 100x100 grid on the detector. Left plot is for camera 1, and right plot is for camera 2.}
    \label{fig:avalanche_gain_plot}
\end{figure}

To gain a better understanding of the pixel-by-pixel gain behavior, our analysis focused on the dataset gathered at a frame rate of 1740 fps, as illustrated in Figure \ref{fig:avalanche_gain_plot}. Across all pixels of camera 1, we observed a linear behaviour, with slope averaging at approximately 0.55. This implies that the applied gain typically represents 55\% of the set gain value. Similarly, in camera 2, a similar trend emerged with pixels exhibiting an average slope of around 0.7, indicating that the applied gain represents 70\% of the set gain value. This discrepancy might be either because the gain calibration against the bias voltage done by the manufacturer has changed, or because we measure it differently.

\subsection{Conversion gain}
To convert the relative digital number output, in the form of ADU, to physical units, a conversion factor is required. The fact that the temporal variance of a photometrically stable signal is proportional to the flux of the signal can be used to calculate this conversion factor. A conversion factor can be inferred from the slope of a plot of the temporal signal variance versus the mean signal. This conversion factor is known as the conversion gain, and the curve plotted is known as the Photon Transfer Curve (PTC).

\[
K \left( \frac{e^-}{\text{ADU}} \right) = \frac{S(\text{ADU})}{\text{Var}(\text{ADU})^2}
\]

\begin{figure}[H]
    \centering
    \begin{subfigure}[b]{0.495\textwidth}
        \centering
        \includegraphics[width=\textwidth]{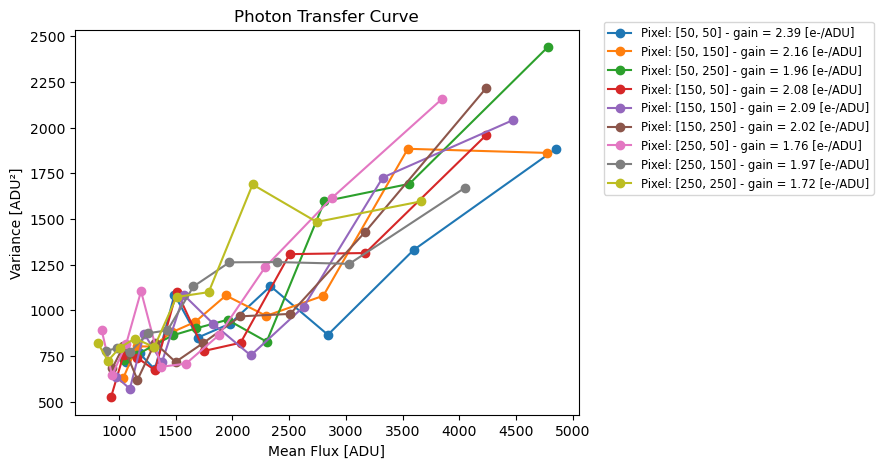} 
        \caption{Camera 1}
        \label{subfig:camera1}
    \end{subfigure}
    \hfill
    \begin{subfigure}[b]{0.495\textwidth}
        \centering
        \includegraphics[width=\textwidth]{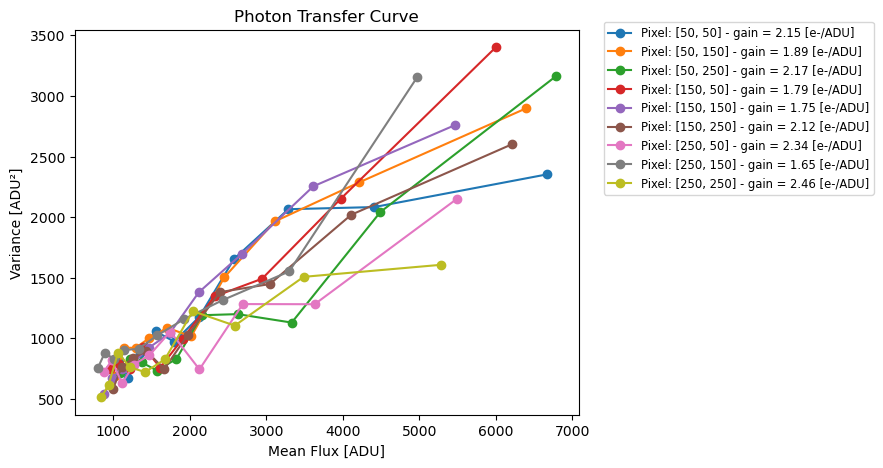} 
        \caption{Camera 2}
        \label{subfig:camera2}
    \end{subfigure}
    \caption{The PTC plotted between mean and variance for various integration times. The frame rate is systematically reduced in 50 fps increments, while the integration time is correspondingly increased as 1/framerate. The pixels included in this plot are chosen based on a 100x100 grid on the detector. Left plot is for camera 1, and right plot is for camera 2.}
    \label{fig:variance_vs_mean}
\end{figure}

\begin{figure}[H]
    \centering
    \begin{subfigure}[b]{0.4\textwidth}
        \centering
        \includegraphics[width=\textwidth]{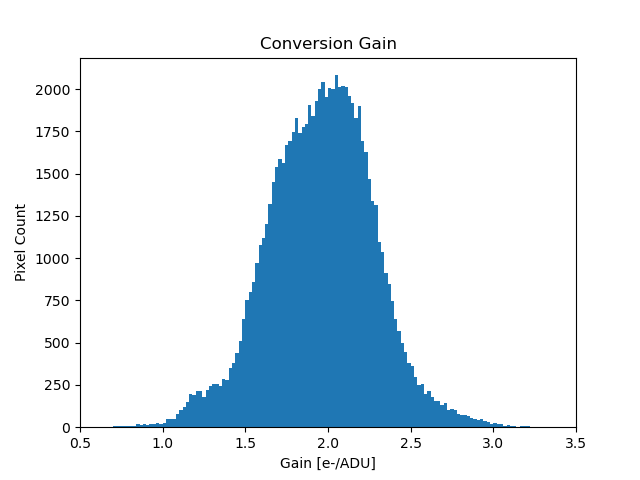} 
        \caption{Camera 1}
        \label{subfig:camera1}
    \end{subfigure}
    \hfill
    \begin{subfigure}[b]{0.4\textwidth}
        \centering
        \includegraphics[width=\textwidth]{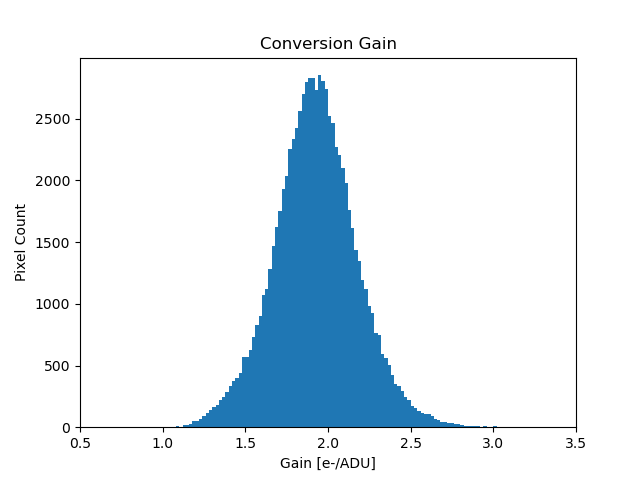} 
        \caption{Camera 2}
        \label{subfig:camera2}
    \end{subfigure}
    \caption{Histograms of conversion gain for all pixels on the detector. Left plot is for camera 1, and right plot is for camera 2.}
    \label{fig:conversion_gain_histogram}
\end{figure}

\begin{figure}[H]
    \centering
    \begin{subfigure}[b]{0.45\textwidth}
        \centering
        \includegraphics[width=\textwidth]{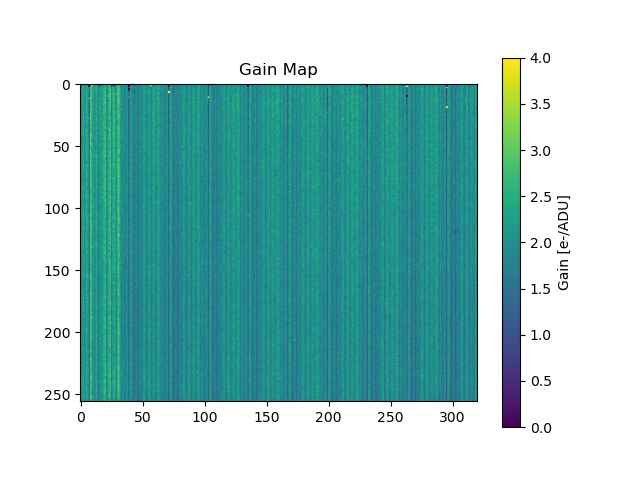} 
        \caption{Camera 1}
        \label{subfig:camera1}
    \end{subfigure}
    \hfill
    \begin{subfigure}[b]{0.45\textwidth}
        \centering
        \includegraphics[width=\textwidth]{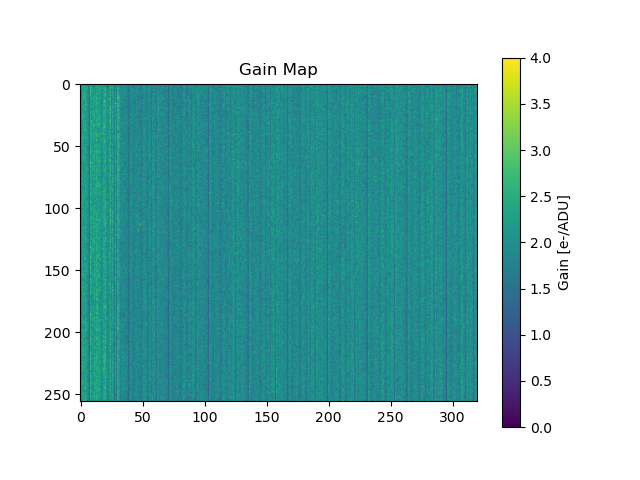} 
        \caption{Camera 2}
        \label{subfig:camera2}
    \end{subfigure}
    \caption{2D map of conversion gain across the detector. Left plot is for camera 1, and right plot is for camera 2.}
    \label{fig:conversion_gain_map}
\end{figure}

In order to determine the conversion gain, flat fields containing 100 frames each were acquired at each integration time with an eAPD-gain of 1, increasing the integration time from 0.5 ms to 10 ms. We observed that the mean signal increases linearly with integration time. Figure \ref{fig:variance_vs_mean} presents the photon transfer curve, where the variance is plotted against the mean signal. The pixel-wise conversion gain is computed by taking the inverse of the slope of the photon transfer curve for each of the pixels. A histogram of conversion gains is shown in Figure \ref{fig:conversion_gain_histogram}. For the camera 1, the median gain is measured to be 2.11 e\textsuperscript{-}/ADU, while the gain value mentioned in test report by FLI is  2.37 [e\textsuperscript{-}/ADU]. For the camera 2, the median gain is measured to be 1.93 e\textsuperscript{-}/ADU, while the gain value mentioned in test report by FLI is  1.76 0[e\textsuperscript{-}/ADU]. To check the spatial variation of conversion, we plot 2D conversion gain maps across the detector as shown in Figure \ref{fig:conversion_gain_map}. We observe that there is a spatial structure that repeats after every 32 pixels, which means the conversion gain varies from one channel to another. We note that the first set of 32 pixels have a higher slightly conversion gain than the rest of the pixels. In the 2D map, dead pixels show a zero gain and hot pixels show a high gain. We also checked the evolution of conversion gain over a period of few hours. The variation in the conversion gain was very small and of the order of 0.02 e\textsuperscript{-}/ADU.

\subsection{Readout noise}
\begin{figure}[htbp]
    \centering
    \begin{subfigure}[b]{0.45\textwidth}
        \centering
        \includegraphics[width=\textwidth]{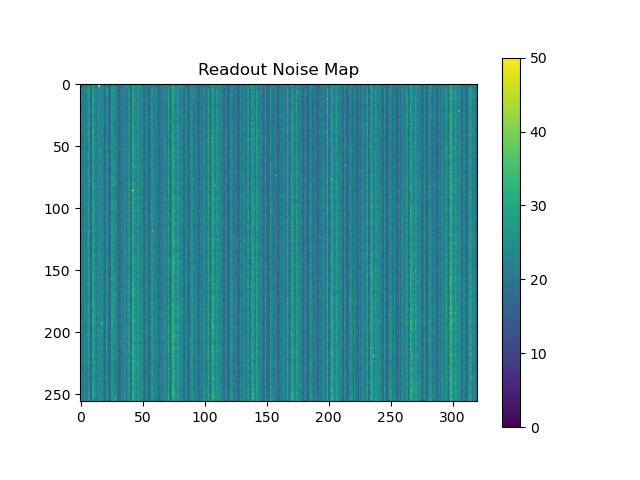} 
        \caption{Camera 1}
        \label{subfig:camera1}
    \end{subfigure}
    \hfill
    \begin{subfigure}[b]{0.45\textwidth}
        \centering
        \includegraphics[width=\textwidth]{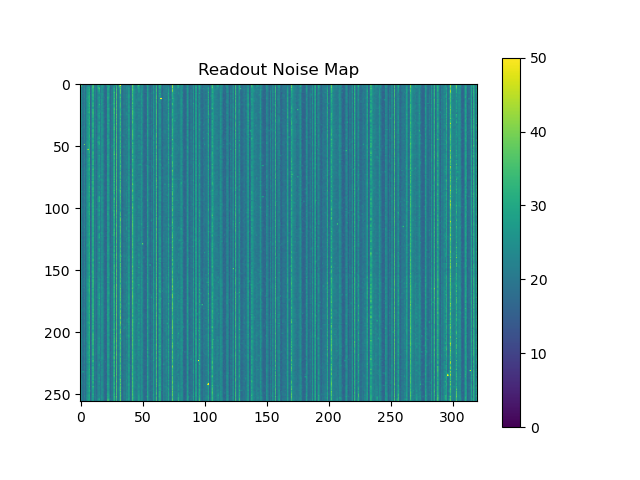} 
        \caption{Camera 2}
        \label{subfig:camera2}
    \end{subfigure}
    \caption{the readout noise map corresponding to the detector operating in CDS mode at its maximum framerate of 1740 fps. The left image corresponds to camera 1, while the right image corresponds to camera 2. The colorbar represents the noise level in ADU.}
    \label{fig:readout_noise_map}
\end{figure}

\begin{figure}[htbp]
    \centering
    \begin{subfigure}[b]{0.45\textwidth}
        \centering
        \includegraphics[width=\textwidth]{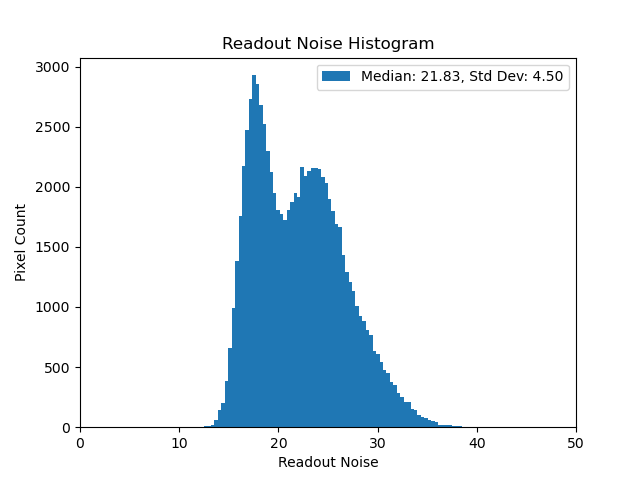} 
        \caption{Camera 1}
        \label{subfig:camera1}
    \end{subfigure}
    \hfill
    \begin{subfigure}[b]{0.45\textwidth}
        \centering
        \includegraphics[width=\textwidth]{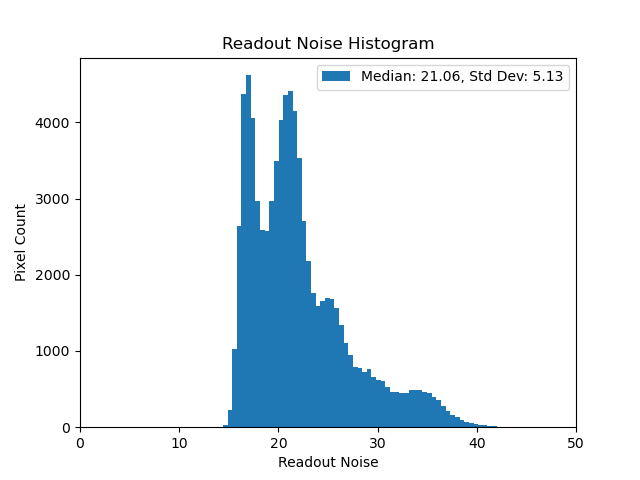} 
        \caption{Camera 2}
        \label{subfig:camera2}
    \end{subfigure}
    \caption{The histogram illustrating the readout noise values for individual pixels. The left image corresponds to camera 1, while the right image corresponds to camera 2.}
    \label{fig:readout_noise_histogram}
\end{figure}

The readout noise is determined by calculating the temporal standard deviation of each pixel across the entire detector frame for a data cube of 100 flat frames acquired at the maximum framerate and at a eAPD-gain of 50.
In Figure \ref{fig:readout_noise_map}, a readout noise map is shown, revealing a similar readout noise pattern in all of the ten subsets, consisting of 32 pixels each, along the readout direction. In Figure \ref{fig:readout_noise_histogram}, the histogram of the readout noise values for individual pixels is plotted. For the camera 1, this histogram shows a bimodal distribution with a median readout noise of 21.83 ADU. The standard deviation of readout noise is 236.40 ADU. However, when the pixels with readout noise higher than 100 ADU are removed, the standard deviation reduces to 4.50 ADU. For the camera 2, the readout noise histogram has a distribution with several peaks. The  median readout noise of 21.06 ADU. The standard deviation of readout noise is 72.67 ADU, which reduces, the standard deviation reduces to 5.13 ADU when pixels with readout noise higher than 100 ADU are removed. The high readout noise pixels are located mostly at the edge of the sensor and thus will have no impact on the data taken. 
We multiply the readout noise value in ADU by the conversion gain and divide it by the eAPD-gain of 50 to obtain the readout noise value in e-. For camera 1, the readout noise is 0.92 e-, whereas for camera 2, it is 0.79 e-.

For a more comprehensive understanding of the distribution of readout noise, we plot the readout noise histograms of all the individual 32 channels. To generate a histogram for a particular channel, all the pixels read by that channel are grouped together. Figure 

\begin{figure}[htbp]
    \centering
    \begin{subfigure}[b]{0.495\textwidth}
        \centering
        \includegraphics[width=\textwidth]{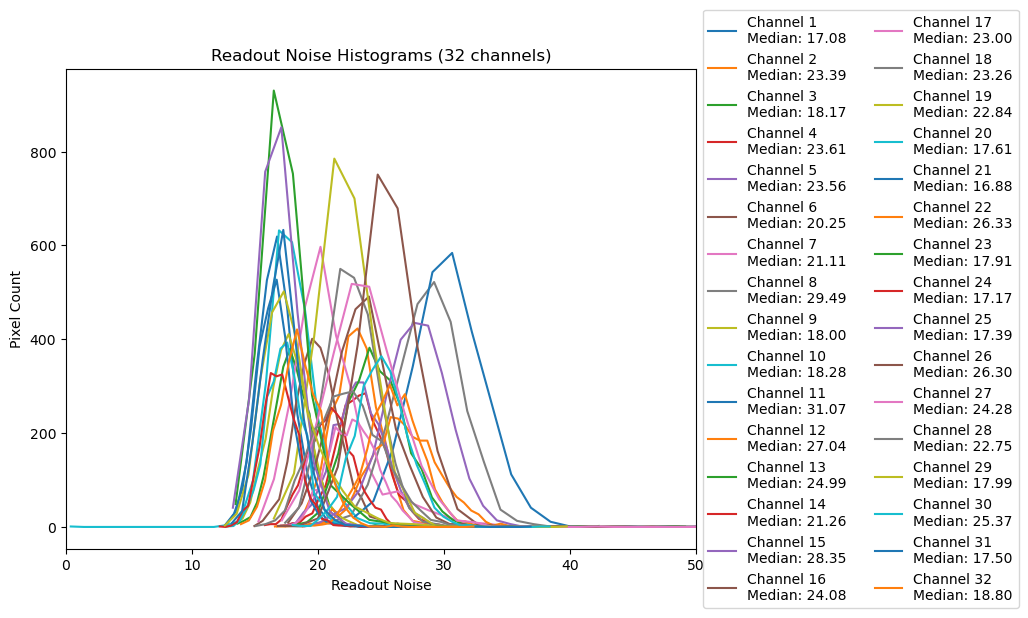} 
        \caption{Camera 1}
        \label{subfig:camera1}
    \end{subfigure}
    \hfill
    \begin{subfigure}[b]{0.495\textwidth}
        \centering
        \includegraphics[width=\textwidth]{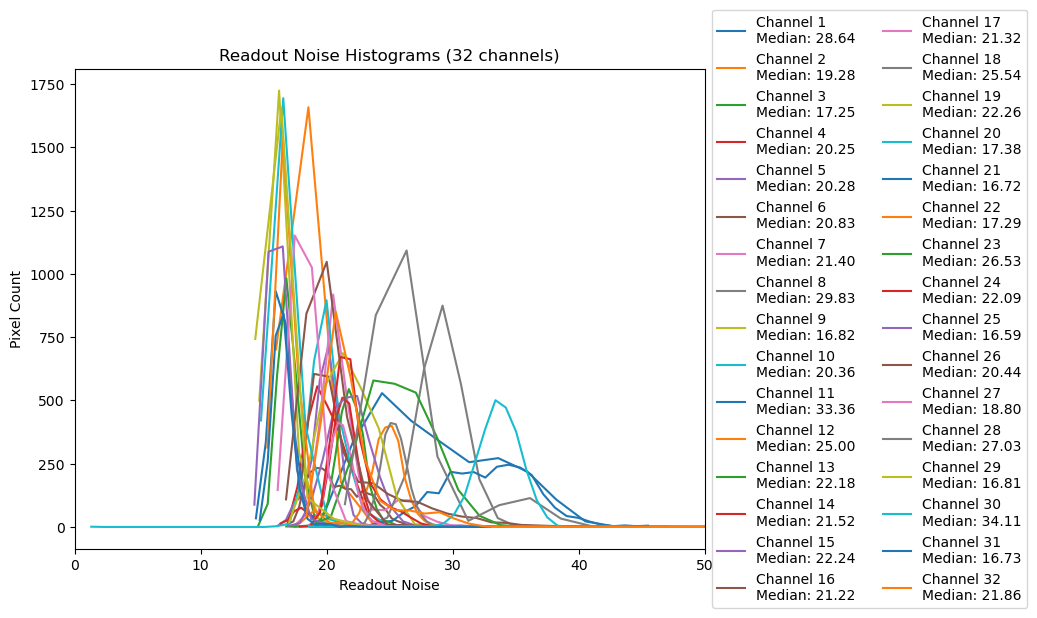} 
        \caption{Camera 2}
        \label{subfig:camera2}
    \end{subfigure}
    \caption{The histograms illustrating the readout noise values for the 32 individual channels. The plot's legend includes information about channel number, the median readout noise for each channel, and the standard deviation of the readout noise. The left plot corresponds to camera 1, while the right plot corresponds to camera 2.}
    \label{fig:readout_noise_histogram_32_channels}
\end{figure}

Figure \ref{fig:readout_noise_histogram_32_channels} displays histograms depicting the distribution of readout noise values for these channels. The readout noise from different channels varies significantly, leading to distinct modes in the distribution. This statistical variability leads to the apparent distribution pattern in the net histogram of readout noise shown in Figure \ref{fig:readout_noise_histogram}.

\subsection{Full well capacity}
The Full well capacity  is a parameter that measures the maximum amount of charge or electrons that a single pixel in the camera can hold before reaching the saturation level. The electrons are collected in a potential well within the pixel until the exposure time ends or the well reaches its full well capacity. 

\begin{figure}[H]
    \centering
    \begin{subfigure}[b]{0.495\textwidth}
        \includegraphics[width=\textwidth]{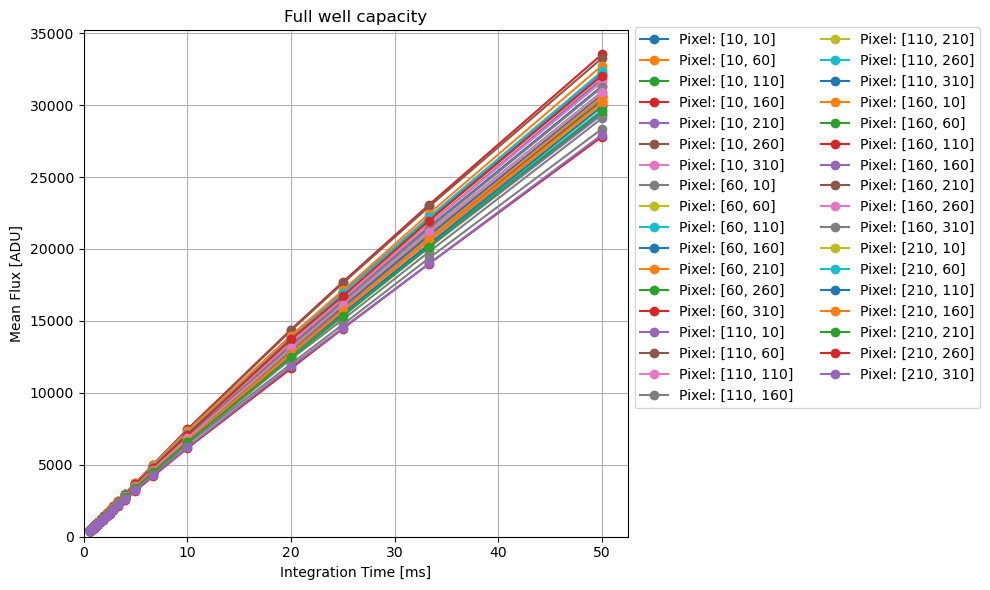} 
        \caption{Camera 1}
        \label{fig:camera 1}
    \end{subfigure}
    \hfill
    \begin{subfigure}[b]{0.495\textwidth}
        \includegraphics[width=\textwidth]{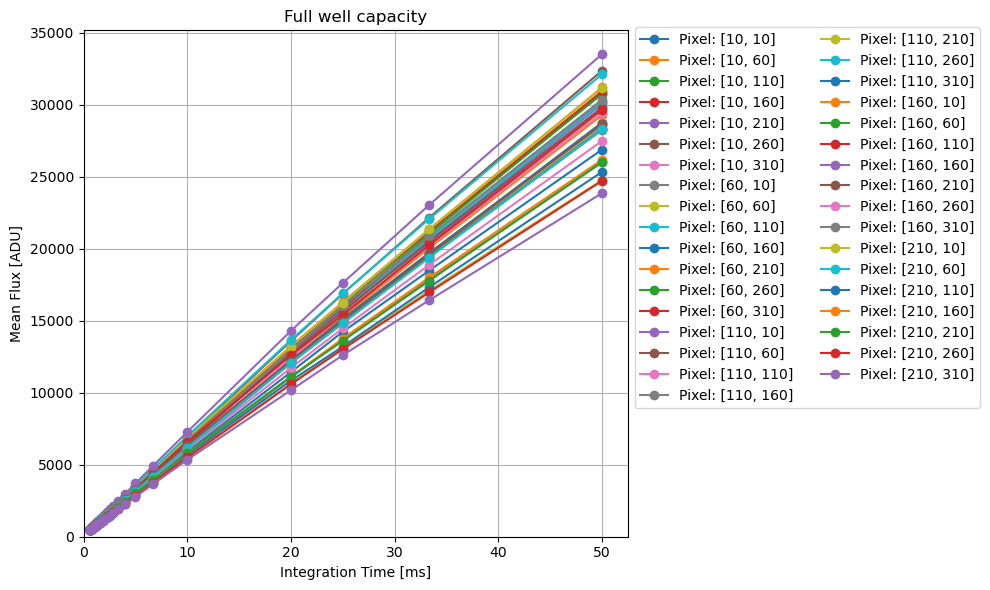} 
        \caption{Camera 2}
        \label{fig:camera 2}
    \end{subfigure}
    
    \caption{The mean flux is plotted as a function of integration time. The frame rate is systematically reduced in 50 fps increments, while the integration time is correspondingly increased as 1/framerate.  The pixels included in this plot are chosen based on a 100x100 grid on the detector. Left plot is for camera 1, and right plot is for camera 2.}
    \label{fig:full_well_capacity_plot}
\end{figure}

To measure the full well capacity, the integration time on the camera was progressively increased till it reached a saturation. The flat fields containing 100 frames each were acquired at each integration time with an eAPD-gain of 1, increasing the integration time in step starting from 0.5 ms.  At integration time of 50 ms, when a pixel detects saturation, the safety feature of the camera kicks in to prevent damaging pixels due to over-saturation and the camera goes in to a safe mode for overillumination. Figure \ref{fig:full_well_capacity_plot} shows the mean flux as a function of integration time. The mean flux exhibits a linear increase as the integration time increases, until it reaches the point of saturation. It is worthy to note that different pixels exhibit slightly varying saturation levels, with each pixel having its own saturation point. We were not able to measure the true saturation points of each of the pixels because the camera doesn't allow to take measurements after the first pixel saturates. We will call the frames taken at 50ms our saturation frames and the flux recorded in the pixels at this 
point their saturation point or full well capacity.

\begin{figure}[htbp]
    \centering
    \begin{subfigure}[b]{0.45\textwidth}
        \centering
        \includegraphics[width=\textwidth]{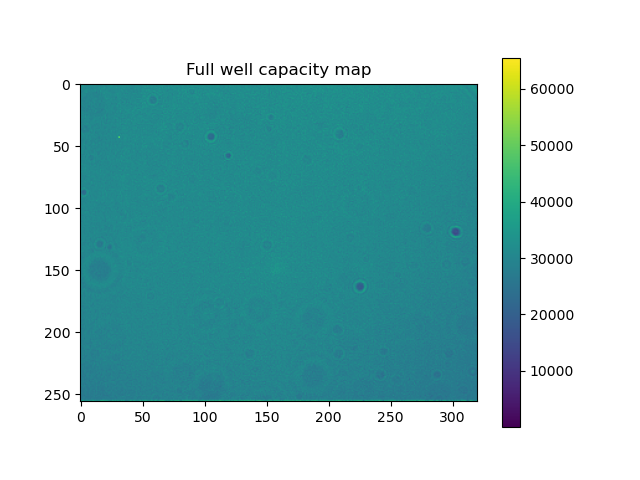} 
        \caption{Camera 1}
        \label{subfig:camera1}
    \end{subfigure}
    \hfill
    \begin{subfigure}[b]{0.45\textwidth}
        \centering
        \includegraphics[width=\textwidth]{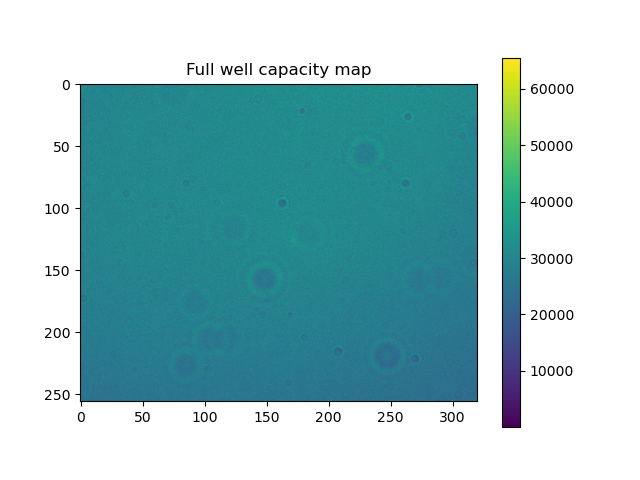} 
        \caption{Camera 2}
        \label{subfig:camera2}
    \end{subfigure}
    \caption{The map of full well capacity maps of the pixels on the detector. The left image corresponds to camera 1, while the right image corresponds to camera 2. The colorbar represents the flux level in ADU.}
    \label{fig:full_well_capacity_map}
\end{figure}

Figure \ref{fig:full_well_capacity_map} shows a map of the full well capacity of all the pixels on the detector or rather the flux recorded in each pixel before camera goes to the safe mode. For the camera 1, the median full well capacity is 30653 ADU, which converts to 64678 e-. While for the camera 2, median full well capacity is 29398 ADU, which converts to 58502 e-. The full-well capacity we measured was quite different from the value of 10000 e- (at e-APD gain of 10) mentioned in FLI test reports, but it was close to the value of 60000 e- in the FLI user manual.

\section{Conclusions}
\noindent

We conducted a characterization of the two CRED One cameras using flat field frames and dark frames acquired through our optical bench setup. Our study included a cosmetic analysis of the detector to identify defective pixels, which showed that there were around 20-30 defective pixels for each of the cameras. The defective pixels we detected were spread across the detector and were not clustered. This means that any region of the detector could be reliably utilized for capturing images from the wavefront sensor. Furthermore, we checked the eAPD-gain calibration of the cameras. We measured the eAPD-gain factor to be quite different from what was claimed by FLI. For camera 1, we found it to be 0.55 and for camera 2, to be 0.70. This discrepancy might be either because the gain calibration against the bias voltage done by the manufacturer has changed, or because we measured it differently. Additionally, we performed a classical photon transfer method analysis to determine the conversion gain of the camera. The values we measured were close to those reported by FLI. However, we observed a spatial variation in conversion across the detector. We also assessed the readout noise at eAPD-gain 1 as well as at higher eAPD-gains. The readout values remained constant in ADUs at different eAPD-gains. To calculate the value in e-, we multiplied it by the conversion gain and divided it by the eAPD-gain. Since FLI had used an eAPD-gain of 50 for their tests, we also reported the values at an eAPD-gain of 50 for our tests. The readout noise values we measured were very close to those measured by FLI, and the values were below 1 e-. Therefore, we could say that we found the camera to be capable of single photon counting. Moreover, we determined the full well capacity of the cameras. The full-well capacity we measured was quite different from the value of 10000 e- mentioned in FLI test reports, but it was close to the value of 60000 e- in the FLI user manual. A summary of our results is shown in the Table~\ref{tab:characterization_test_results}.

\renewcommand{\arraystretch}{2} 

\begin{table}[h!]
\centering
\caption{Characterization test results}
\vspace{1em}
\label{tab:characterization_test_results}
\begin{tabularx}{\textwidth}{|X|X|X|X|X|X|}
\hline
\multirow{2}{*}{\makecell{\textbf{Specification}}} & \multirow{2}{*}{\textbf{Unit}} & \multicolumn{2}{c|}{\textbf{Camera 1}} & \multicolumn{2}{c|}{\textbf{Camera 2}} \\
\cline{3-6}
& & \textbf{FLI tests} & \textbf{Our tests} & \textbf{FLI tests} & \textbf{Our tests} \\
\hline
Array size & pixels & 320x256 & 320x256 & 320x256 & 320x256 \\
\hline
Maximum speed at full-frame single-read & fps & 3500 & 3502 & 3500 & 3504 \\
\hline
Bias level & ADU & -- & 214 & -- & 314 \\
\hline
Defective pixels & pixels & 0 & 20 (at eAPD-gain of 108) & 1 & 29 (at eAPD-gain of 96)\\
\hline
Avalanche gain (eAPD-gain) & -- & 108 & 59.4 & 96 & 67.2 \\
\hline
Conversion gain & e-/ADU & 2.37 & 2.11 & 1.76 & 1.96 \\
\hline
Readout noise CDS 1720 fps eAPD-gain x50 & e-/pixel/frame & 0.90 & 0.92 & 0.72 & 0.79 \\
\hline 
Full well capacity & e- & 10000 (at eAPD-gain of 10) & 64678 (at eAPD-gain of 1) & 10000 (at eAPD-gain of 10) & 58502 (at eAPD-gain of 1) \\
\hline
\end{tabularx}
\end{table}

\acknowledgments     
 
This work has been carried out within the framework of the National Centre of Competence in Research PlanetS supported by the Swiss National Science Foundation under grants 51NF40\_182901 and 51NF40\_205606. The RISTRETTO project was partially funded through the SNSF FLARE programme for large infrastructures under grants 20FL21\_173604 and 20FL20\_186177. The authors acknowledge the financial support of the SNSF.

\vspace{2\baselineskip}
\bibliographystyle{spiebib}
\bibliography{cred_biblio}

\begin{thebibliography}{1}

\bibitem{finger2014saphira}
G.~Finger, I.~Baker, D.~Alvarez, D.~Ives, L.~Mehrgan, M.~Meyer, J.~Stegmeier, and H.~J. Weller, ``Saphira detector for infrared wavefront sensing,'' in {\em Adaptive Optics Systems IV},   {\bf 9148}, pp.~427--442, SPIE, 2014.

\bibitem{finger2010development}
G.~Finger, I.~Baker, R.~Dorn, S.~Eschbaumer, D.~Ives, L.~Mehrgan, M.~Meyer, and J.~Stegmeier, ``Development of high-speed, low-noise nir hgcdte avalanche photodiode arrays for adaptive optics and interferometry,'' in {\em High Energy, Optical, and Infrared Detectors for Astronomy IV},   {\bf 7742}, pp.~471--484, SPIE, 2010.

\bibitem{greffe2016c}
T.~Greffe, P.~Feautrier, J.-L. Gach, E.~Stadler, F.~Clop, S.~Lemarchand, D.~Boutolleau, and I.~Baker, ``C-red one: the infrared camera using the saphira e-apd detector,'' in {\em Optical and Infrared Interferometry and Imaging V},   {\bf 9907}, pp.~634--642, SPIE, 2016.

\bibitem{chazelas2020ristretto}
B.~Chazelas, C.~Lovis, N.~Blind, J.~K{\"u}hn, L.~Genolet, I.~Hughes, M.~Turbet, J.~Hagelberg, N.~Restori, M.~Kasper, {\em et~al.}, ``Ristretto: a pathfinder instrument for exoplanet atmosphere characterization,'' in {\em Adaptive Optics Systems VII},   {\bf 11448}, pp.~1393--1401, SPIE, 2020.

\bibitem{lovis2022ristretto}
C.~Lovis, N.~Blind, B.~Chazelas, J.~G. K{\"u}hn, L.~Genolet, I.~Hughes, M.~Sordet, R.~Schnell, M.~Turbet, T.~Fusco, {\em et~al.}, ``Ristretto: high-resolution spectroscopy at the diffraction limit of the vlt,'' in {\em Ground-based and airborne instrumentation for astronomy IX},   {\bf 12184}, pp.~590--598, SPIE, 2022.

\bibitem{stadler2022saxo+}
E.~Stadler, E.~Diolaiti, L.~Schreiber, F.~Cortecchia, M.~Lombini, M.~Loupias, Y.~Magnard, A.~De~Rosa, G.~Malaguti, D.~Maurel, {\em et~al.}, ``Saxo+, a second-stage adaptive optics for sphere on vlt: optical and mechanical design concept,'' in {\em Adaptive Optics Systems VIII},   {\bf 12185}, pp.~1318--1328, SPIE, 2022.

\bibitem{gach2016c}
J.-L. Gach, P.~Feautrier, E.~Stadler, T.~Greffe, F.~Clop, S.~Lemarchand, T.~Carmignani, D.~Boutolleau, and I.~Baker, ``C-red one: ultra-high speed wavefront sensing in the infrared made possible,'' in {\em Adaptive Optics Systems V},   {\bf 9909}, pp.~374--383, SPIE, 2016.

\end{thebibliography}

\end{document}